\documentclass{PoS}
\usepackage{amsmath}

\title{Error reduction technique using covariant approximation and application to nucleon form factor}

\ShortTitle{Error reduction technique using covariant approximation and application to nucleon form factor}

\author{Thomas Blum\\
        Physics Department, University of Connecticut, Storrs, CT 06269-3046, USA\\
        RIKEN-BNL Research Center, Brookhaven National Laboratory, Upton, NY 11973, USA\\
        E-mail: \email{tblum@phys.uconn.edu}}

\author{Taku Izubuchi\\
        Brookhaven National Laboratory, Upton, NY 11973, USA\\
        RIKEN-BNL Research Center, Brookhaven National Laboratory, Upton, NY 11973, USA\\
        E-mail: \email{izubuchi@quark.phy.bnl.gov}}

\author{\speaker{Eigo Shintani}\\
        RIKEN-BNL Research Center, Brookhaven National Laboratory, Upton, NY 11973, USA\\
        E-mail: \email{shintani@riken.jp}}

\abstract{We demonstrate the new class of variance reduction techniques for 
hadron propagator and nucleon isovector form factor in the realistic lattice 
of $N_f=2+1$ domain-wall fermion.
All-mode averaging (AMA) is one of the powerful tools to 
reduce the statistical noise effectively for wider varieties of 
observables compared to existing techniques such as low-mode averaging (LMA). 
We adopt this technique to hadron two-point functions and 
three-point functions, and compare with LMA and traditional 
source-shift method in the same ensembles.
We observe AMA is  much more cost effective in reducing statistical error
for these observables.
}
\FullConference{The 30 International Symposium on Lattice Field Theory - Lattice 2012,\\
		June 24-29, 2012\\
		Cairns, Australia}
\begin{document}
\section{Introduction}
In order to precisely evaluate non-perturbative quantities in lattice calculation, 
the reducing noise-to-signal ratio is one of 
the most important tasks especially for 
nucleon electric dipole moment~\cite{Shintani:2005xg,Berruto:2005hg}, 
the hadronic contribution to muon anomalous 
magnetic moment~\cite{Aubin:2012me, Blum:Lat2012}, 
nucleon form factors and structure functions~\cite{Yamazaki:2009zq},
$\eta'$ meson mass and mixing angle~\cite{Christ:2010dd}  
and so on. 
We consider new strategies to effectively increase statistics 
without generating the new gauge configurations.

Traditionally translational symmetry on the lattice has been used to 
increase the statistics of correlation function (correlator) 
of hadron interpolating operator(s). 
Since correlators for different source locations 
with the same relative separations between operators are exactly 
same in the infinite statistics limit
for translational invariant action, 
the average over several source locations 
can be regarded as several times more statistical samples 
if correlation among correlators with different source is negligible. 
In this case, however, the additional computation of conjugate gradient 
(CG) at each source locations is needed. 
Low-mode-averaging (LMA)~\cite{Giusti:2002sm,Giusti:2004yp,DeGrand:2004qw} 
takes  advantage of low-(eigen)mode for its dominant
observables {\it e.g.} pseudoscalar correlator
\cite{Giusti:2002sm,Giusti:2004yp,DeGrand:2004qw,DeGrand:2005vb,Luscher:2007se,Fukaya:2007fb},
to remove the cost of additional computation of CG.
However for the other observables, nucleon propagator or resonance state
and heavy mesons, 
\cite{Giusti:2005sx,Li:2010pw,Bali:2010se} reported that 
the statistical error reduction in LMA 
is less significant than pseudoscalar correlator.

Here we examine the different variance reduction techniques from LMA
for nucleon correlator and its three-point function.
All-mode averaging (AMA) proposed in \cite{Blum:2012uh} 
reduces variance from {\it all}-modes, without introducing bias, 
and thus should serve as a 
powerful technique to precisely evaluate the observables 
including highly composite contribution coming from {\it all}-modes.
In this proceedings, we explain this idea and show the numerical results 
of hadron spectroscopy and isovector form factors of nucleon 
in realistic lattice setup.

\section{Covariant approximation averaging}
The observable $\mathcal O$ is  calculated over 
gauge ensemble $\{U_1,U_2,\cdots,U_{N_{\rm conf}}\}$ to obtain the 
ensemble average;
\begin{equation}
  \langle \mathcal O\rangle = \frac{1}{N_{\rm conf}}\sum_{i=1}^{N_{\rm conf}} \mathcal O[U_i] 
  + O(N_{\rm conf}^{-1/2}),
\end{equation}
where the second term denotes the uncertainty 
due to finite number of available gauge configurations.
Under a  transformation $g\in G$, where $G$ is 
a set of symmetry transformations on the lattice,
the link variable is transformed as $U \rightarrow U^g$. 
Ensemble average of an observable ${\cal O}$ should be 
covariant,
\begin{equation}
 \langle\mathcal O^g[U]\rangle = \langle\mathcal O[U^g]\rangle,
\label{eq:covariance}
\end{equation}
in the infinite statistics $N_{\rm conf}\rightarrow\infty$.
${\cal O}^g$ is the transformed observables.

Here we introduce the approximation $\mathcal O^{\rm (appx)}$ 
which fulfills the following condition:\\
{\bf Appx-1}: 
\begin{equation}
  r = \frac{\langle\Delta\mathcal O\Delta\mathcal O^{(\rm appx)}\rangle}
      {\sqrt{\langle(\Delta\mathcal O)^2\rangle\langle(\Delta\mathcal O^{(\rm appx)})^2\rangle}}
    \simeq 1,\quad
  \langle(\Delta\mathcal O)^2\rangle \simeq \langle(\Delta\mathcal O^{(\rm appx)})^2\rangle,
  \label{eq:corr}
\end{equation}
with $\Delta X = X - \langle X\rangle$. $r$ is the correlation between
$\mathcal O$ and $\mathcal O^{\rm (appx)}$.\\
{\bf Appx-2}: The computational cost of $\mathcal O^{(\rm appx)}$ is much smaller than original 
              $\mathcal O$.\\
{\bf Appx-3}: Covariance of approximation : 
$\langle\mathcal O^{\rm appx}[U^g]\rangle = \langle\mathcal O^{ {(\rm appx)}\,g }[U]\rangle$.\\
Using $\mathcal O^{\rm (appx)}$  we construct improved estimator;
\begin{equation}
  \mathcal O^{\rm (imp)} = \mathcal O^{\rm (rest)} + \mathcal O_G^{\rm (appx)},\quad
  \mathcal O^{\rm (rest)} = \mathcal O - \mathcal O^{(\rm appx)},\quad
  \mathcal O_G^{\rm (appx)} = N_G^{-1}\sum_{g\in G}\mathcal O^{ {\rm (appx)}\,g}.
  \label{eq:imp}
\end{equation}
The first and the second conditions are to reduce  statistical error 
of ${\cal O^{\rm (rest)}}$ and the computational cost respectively 
at fixed $N_{\rm conf}$, 
while the third one is to avoid the bias : it leads to 
$\langle\mathcal O^{(\rm imp)}\rangle = \langle\mathcal O\rangle$.
When we perform $N_G$ times measurements for $\mathcal O^{\rm (appx)}$ after transformation $g\in G$, 
for instance, shifting source locations, 
the statistical error of improved estimator, $\Delta_{(\rm imp)}$, will 
be reduced to
\begin{equation}
  \Delta_{(\rm imp)} \simeq \Delta\sqrt{2(1-r)+N_G^{-1}},
  \label{eq:error}
\end{equation}
compared with original error $\Delta$ 
ignoring the correlation among  $\mathcal O^{\rm (appx)\,g}$ with different $g$'s. 
In the case of strong correlation, $r\simeq 1$, in {\bf Appx-1}, 
$\Delta_{(\rm imp)}$ becomes nearly $N_G^{-1/2}$ times smaller than $\Delta$.
If Cost($\mathcal O_G^{\rm (appx)}$) is significantly cheaper than
$N_G\times$Cost($\mathcal O$) ({\bf Appx-2}), 
total cost of $\mathcal O^{\rm (imp)}$ for a fixed size of error is reduced
by a factor 
$\sim \text{Cost}({\cal O^\text{(appx)}})/\text{Cost}({\cal O}) + N_G^{-1}$
(the last factor comes from computation of $\mathcal O$ in $\mathcal O^{(\rm rest)}$).
The above improved estimator defines
{\it covariant approximation averaging (CAA)}.

LMA~\cite{Giusti:2004yp,DeGrand:2004qw} is one of  CAA; 
$\mathcal O^{\rm (appx)}$ consists of low-mode, and $g$ is a
shift of source location. 
In LMA, $\mathcal O^{\rm (appx)}$ is correlator constructed 
from a few low-modes of Hermitian Dirac operator $H(x,y)$ 
(or its even-odd preconditioned counterpart),
where we only present formula for the point source case for simplicity,
\begin{equation}
  S^{\rm (low)}(x,y) = \sum_{k=1}^{N_\lambda} \lambda_k^{-1}\psi_k(x)\psi^\dag_k(y),\quad
  \mathcal O_G^{\rm (LMA)} = \frac{1}{N_G}\sum_{g\in G}\mathcal O(S^{\rm (low)\,g}),
  \label{eq:LMA}
\end{equation}
with eigenmode $\psi_k$ and eigenvalue $\lambda_k$ in $\sum_y H(x,y)\psi_k(y)=\lambda_k\psi_k(x)$.
{\bf Appx-3} is satisfied since $\psi_k(x)$ respects the translational symmetry. 
Cost of ${\cal O}^\text{(appx)}$ includes
the I/O for eigenmode, outer products of eigenvectors in Eq.(\ref{eq:LMA}),
and the cost to contract quark propagators for $\mathcal O(S^{\rm (low)})$,
which is typically small cost ({\bf Appx-2}). 
{\bf Appx-2,3} are satisfied for most of observables in LMA, 
however, {\bf Appx-1} strongly depends on ${\cal O}$ and
size of $N_\lambda \sim O(100)$. 

We proposed another example of CAA, which we call as {\it all-mode averaging (AMA)} in \cite{Blum:2012uh}.
Using the sloppy CG~\cite{Bali:2009hu} combined with low-mode deflation
({\it e.g}  \cite{Luscher:2007se})
in which the stopping condition $\varepsilon$ of CG is 
made loose as $\varepsilon_{\rm AMA}<10^{-3}$--$10^{-4}$ 
(or fix the number of CG iterations to some small number
\footnote{
With the fixed stopping condition of CG in
the approximation, $S^{\rm (all)}$, 
a bias will be introduced due to the finite
precision (64 bits arithmetic in our case) breaking {\bf Appx-3}
in theoretically very small provability. 
This bias could be avoided by fixing the iteration number to a constant as
pointed out by M.L\"uscher and S. Hashimoto independently.  
In this proceedings, however, we checked that this bias is undetectable
by comparing results of AMA and LMA, latter of which 
does not have such bias: for pion propagator they are identical 
within 1-$\sigma$ (0.1\%) error, as shown in Figure \ref{fig:Npiv}. 
}), 
the approximation is given by
\begin{equation}
  S^{\rm (all)}(x,y) = \sum_{k=1}^{N_\lambda} \lambda_k^{-1}\psi_k(x)\psi^\dag_k(y) 
  + f_\varepsilon(H(x,y)),\quad
  \mathcal O_G^{\rm (AMA)} = \frac{1}{N_G}\sum_{g\in G}\mathcal O(S^{\rm (all)\,g}),
  \label{eq:AMA}
\end{equation}
where $f_\varepsilon$ denotes the polynomial function of $H$ implicitly 
created in CG process to approximate inverse : $f_\varepsilon(\lambda)\sim 1/\lambda$ for $\lambda_N < \lambda < \lambda_\text{max}$. 
AMA has advantage that $S^{\rm (all)}(x,y)$ takes account of not only 
low-mode contribution but also (approximately) all-mode contribution 
which is controlled by the two parameters $N_\lambda$ and $\varepsilon_{\rm AMA}$.
AMA also fulfills the above three conditions 
({\bf Appx-1}--{\bf Appx-3}) for a much wider class of observables than LMA.

\section{Numerical results}
We use the $N_f=2+1$ domain-wall fermion (DWF) configurations 
generated by RBC/UKQCD collaboration in 24$^3\times$64 lattice 
at $\beta=2.13$ Iwasaki gauge action \cite{Aoki:2010dy}. 
CG algorithm with four dimensional even-odd preconditioning was used 
to compute quark propagators at quark mass $m=0.01, 0.005$, 
and  5th dimension size is $L_s=16$.
To calculate eigenmode of 
Hermitian even-odd preconditioned kernel of DWF operator in $10^{-8}$ accuracy, 
we implement the implicitly restarted Lanczos algorithm 
with Chebychev polynomial acceleration \cite{Neff:2001zr}.
Note that in use of even-odd bases, one needs to choose
the four dimensional shift vector of source point to meet {\bf Appx-3}
(even steps in four directions are sufficient).

In this proceedings, LMA/AMA estimator is obtained 
from ${\cal O}_G^{\rm (appx)}$ with $N_G=32$ different source locations  
separated by every 12 for spatial direction and 16 for temporal direction;
(0,0,0,0), (12,0,0,0), (12,12,0,0),$\cdots$,(12,12,12,48) in  lattice unit. 
(0,0,0,0) is the original source location for ${\cal O}$.
Stopping condition $\varepsilon$ 
for original observable $\mathcal O$, and  $\varepsilon_{\rm AMA}$ 
for the sloppy CG in AMA, are defined 
as $||Hx-b||/||b||<\varepsilon,\varepsilon_{\rm AMA}$ 
with the even-site source vector $b$ and the even-site solution vector $x$ 
(see also Table \ref{tab:param}).
Note that the number of CG iterations is for the case 
of deflated CG with $N_\lambda$ low-modes.
To compare the performance, we set the Gaussian-type smearing source 
with parameters taken from \cite{Yamazaki:2009zq}.
In Ref.\cite{Yamazaki:2009zq} they measured
three- and two-point functions for 
four source locations in temporal direction to extract
the nucleon isovector form factor (and axial charge), 
and thus  $4\times N_{\rm conf}$ samples were accumulated.
For $m=0.005$, they have shown the results
with quark sources set in two timeslices per one CG (double source method), 
by which effective samples are doubled.
Furthermore their nucleon source was non-relativistic definition 
(2 spinor sources rather than 4).
In our cost analysis below, we will correct these two factors
for fair comparison.

\begin{table}
\begin{center}
\caption{Parameters in LMA/AMA. Ranges of CG iteration numbers 
in each ensembles are shown.}\label{tab:param}
\begin{tabular}{cccccccc}
\hline\hline
$m$ & $N_{\rm conf}$ & $N_G$ & $N_\lambda$ & $\varepsilon$ & CG iter. & $\varepsilon_{\rm AMA}$ 
    & CG iter.(AMA)\\
\hline
0.005 & 380 & 32 & 400 & $10^{-8}$ & 350--360 & $3\times 10^{-3}$ & 70--90\\
0.01  & 257 & 32 & 180 & $10^{-8}$ & 600--630 & $3\times 10^{-3}$ & 90--130\\
\hline\hline
\end{tabular}
\end{center}
\end{table}

\subsection{Two-point function}
\begin{figure}
\begin{center}
\includegraphics[width=130mm]{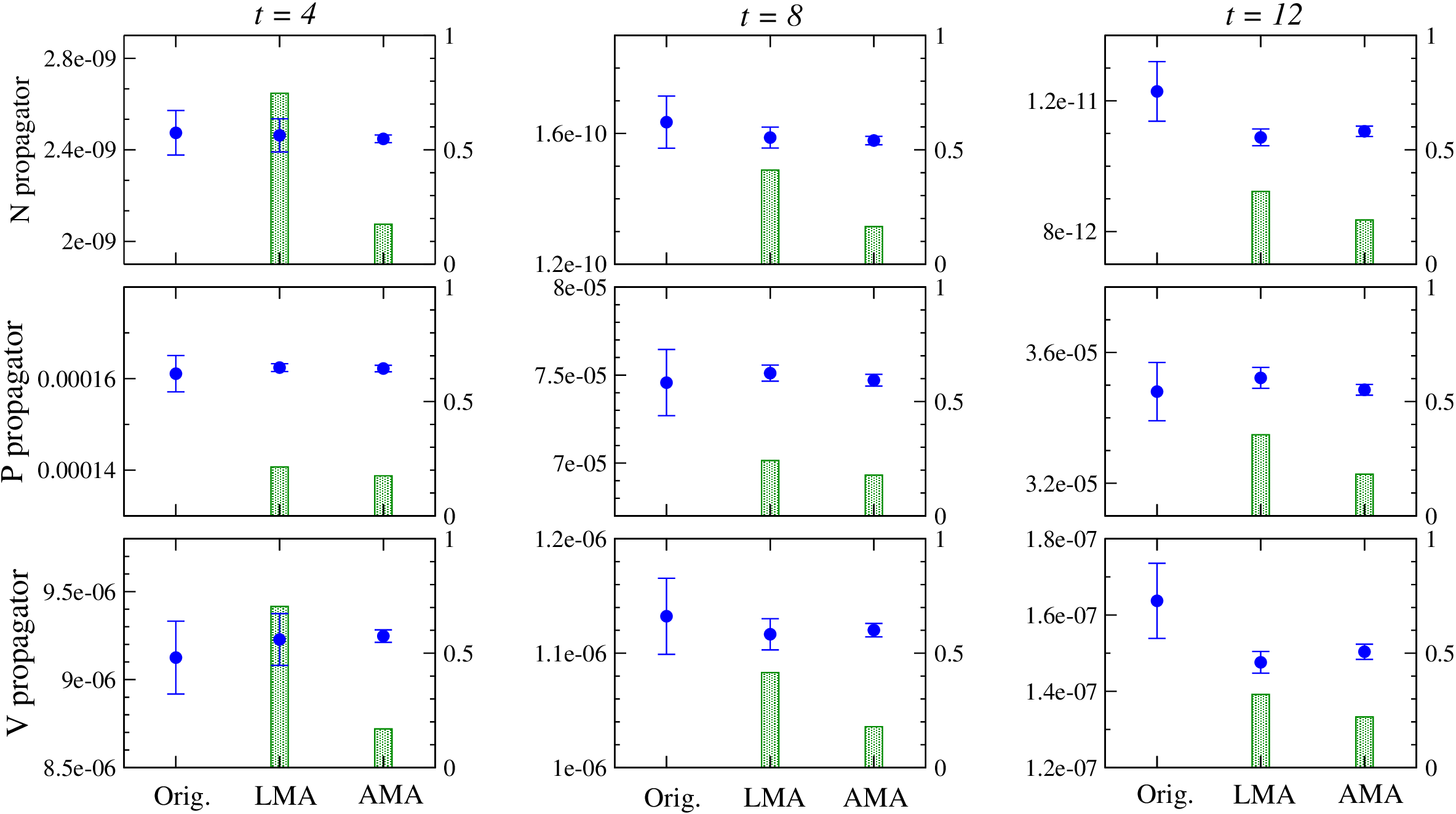}
\caption{The comparison between LMA/AMA and original analysis for nucleon (N), 
pseudoscalar (P) and vector (V) meson propagator at different time-slices $t=4,8,12$.
The colored bar indicates the ratio of relative error between 
original and LMA/AMA. This is the case at $m=0.005$.}\label{fig:Npiv}
\end{center}
\end{figure}

In Figure \ref{fig:Npiv}, we compare  
nucleon ($N$), pseudoscalar ($P$) and vector ($V$) meson correlator
on three different time separations for original and LMA/AMA analysis. 
As mentioned before, LMA suppresses fluctuation from 
low-modes of Dirac matrix in Eq.(\ref{eq:LMA}) by averaging over $N_G$
source locations of low-modes, 
and thus the error reduction of LMA is significant for 
larger time separations. 
On the other hand, since AMA approximately averages contribution 
from all modes, AMA is expected to suppress the fluctuation 
of observables in both short and long distances.
The above expectation is clearly seen in Figure \ref{fig:Npiv}; 
as a time-separation changes from $t=12$ to $t=4$, the error reduction ratio
(green bar) is degraded in LMA significantly, while, for AMA, 
error reduction rate remains close to the ideal magnitude, 
$1/N_G^{1/2}\simeq0.18$, and always smaller than that of LMA, 
for every channels and distances.
Error reduction for $P$ channel is less
different between LMA and AMA. This is likely because the single pion is
mostly dominated by the low-modes, also discussed
in \cite{Giusti:2002sm,Giusti:2004yp,DeGrand:2004qw}.

\begin{figure}
\begin{center}
\includegraphics[width=100mm]{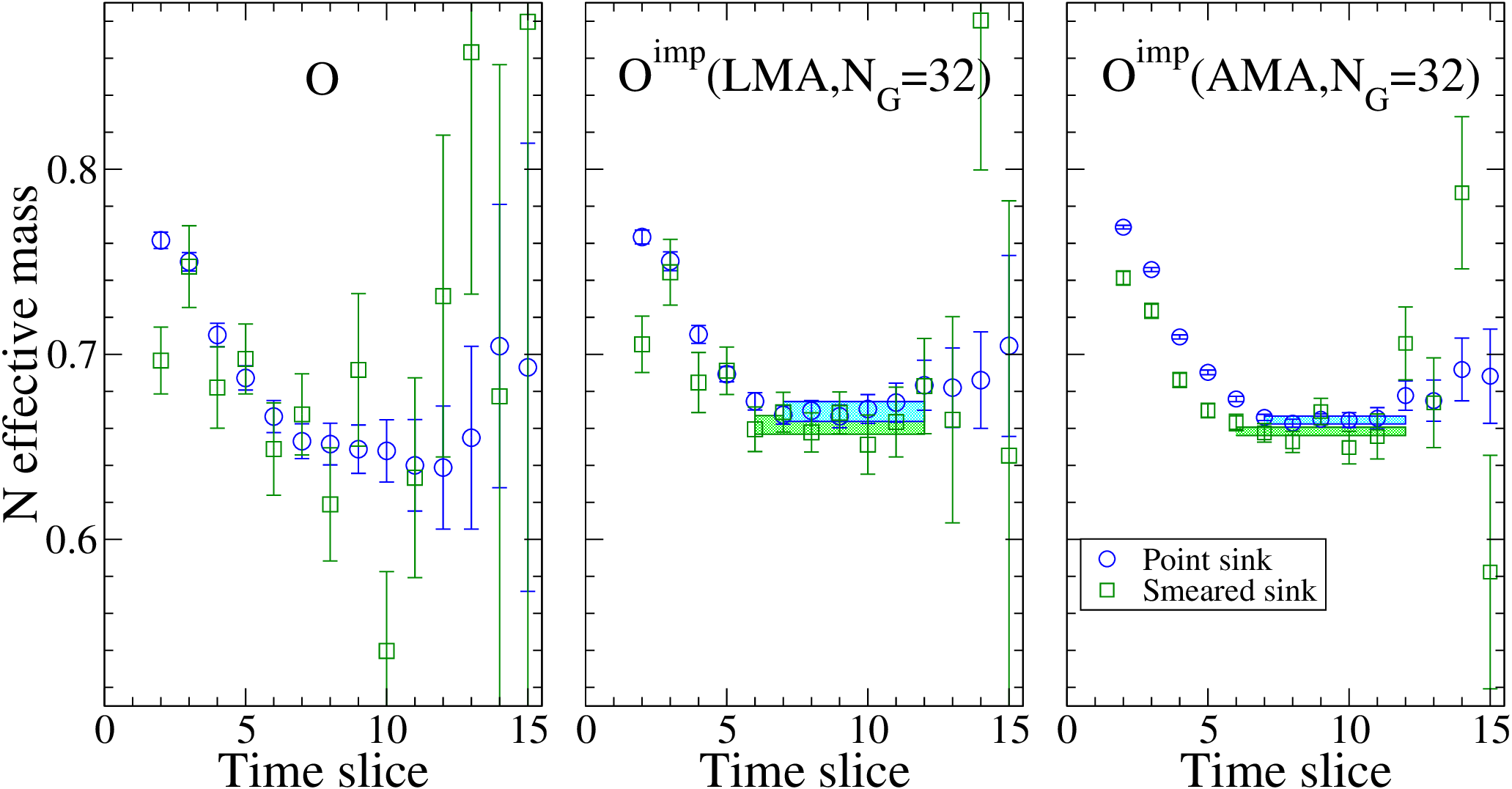}
\caption{Effective mass plot of nucleon correlator in original, LMA/AMA analysis
with point sink and Gaussian smeared sink. The colored bound shows the statistical 
error when globally fitting the propagator.}\label{fig:effm}
\end{center}
\end{figure}

In Figure \ref{fig:effm}, the effective mass of nucleon in AMA
becomes flatter for both point and Gaussian smeared sinks compared to those of LMA. 
In Table \ref{tab:cost} we see that the precision of nucleon mass 
in AMA is higher than previous study \cite{Yamazaki:2009zq} while 
the computational cost is roughly less by 1/4 times. 
The detailed comparison between them including computational time
of low-mode is discussed in \cite{Blum:2012uh}.

\begin{table}
\begin{center}
\caption{The nucleon mass obtained by global fitting of N correlator with 
Gaussian smeared sink. We use GeV unit. 
Last two columns are the gain for AMA with and without deflation, 
which are inverse of reductions of costs to achieve the same size of errors.
For Gain$_{\rm w/\,def}$, the cost for computing eigenvectors are ignored
while Gain$_{\rm w/o\,def}$ reflects its cost, see \cite{Blum:2012uh}
for more detail breakup of cots.
Costs of TY \cite{Yamazaki:2009zq} are estimated
in case of the relativistic source on a single time-slice per a CG for
the fair comparison; 1864 meas. ($m=0.005$) and 1424 meas. in $m=0.01$. 
} 
\label{tab:cost}
\begin{tabular}{ccccccccc}
\hline\hline
$m_\pi$ & $m_N$(Orig) & $m_N$(LMA) & $m_N$(AMA) & $m_N$(TY) 
       & Gain$_{\rm w/\,def}$ & Gain$_{\rm w/o\,def}$\\
\hline
0.33 & 1.124(22) & 1.145(9) & 1.139(4)  & 1.148(10) & 4 & 15 \\
0.42 & 1.221(17) & 1.219(11) & 1.233(4) & 1.217(9)  & 4 & 5 \\
\hline\hline
\end{tabular}
\end{center}
\end{table}

\subsection{Nucleon isovector form factor}
\begin{figure}
\begin{center}
\includegraphics[width=71mm]{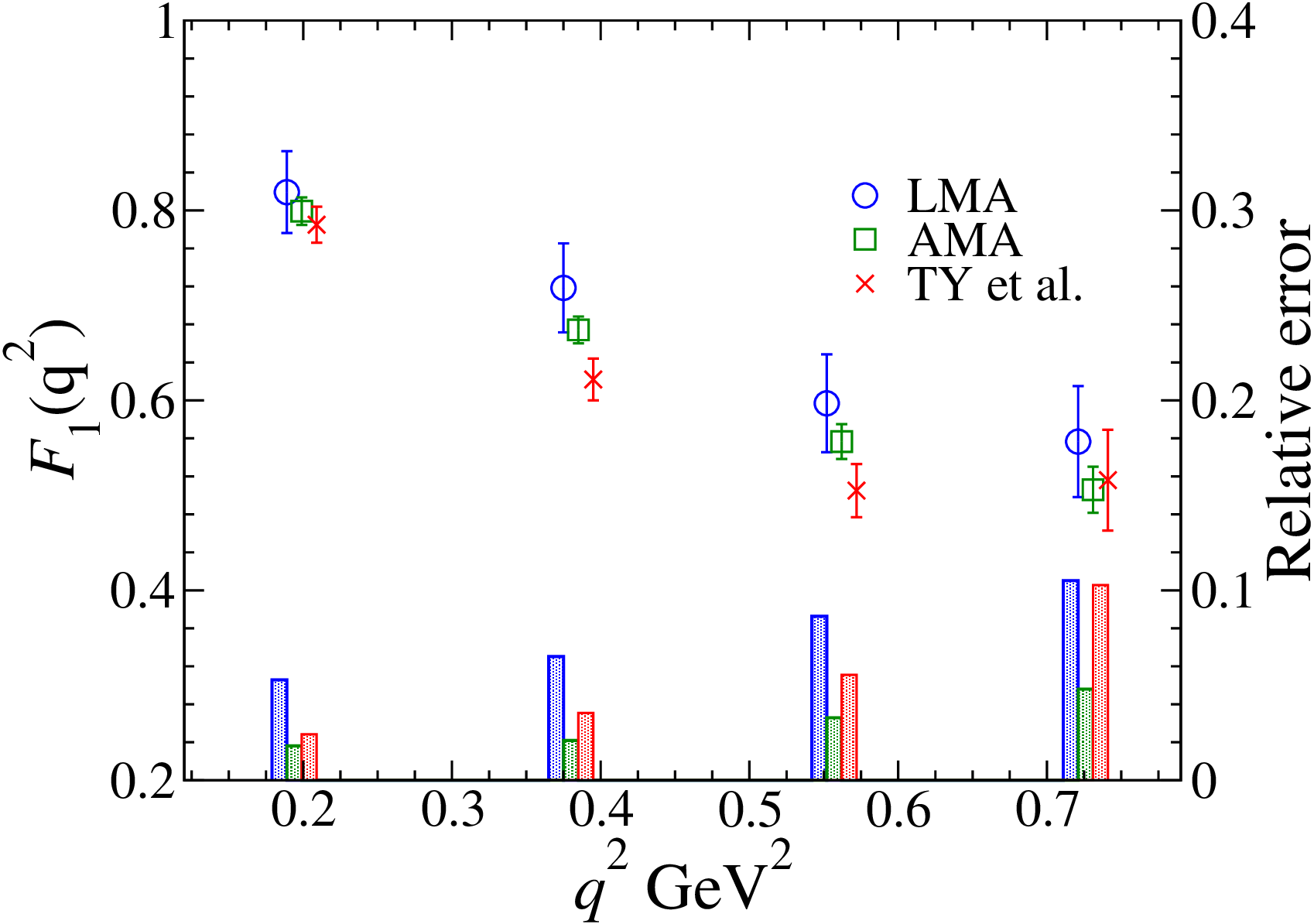}
\includegraphics[width=70mm]{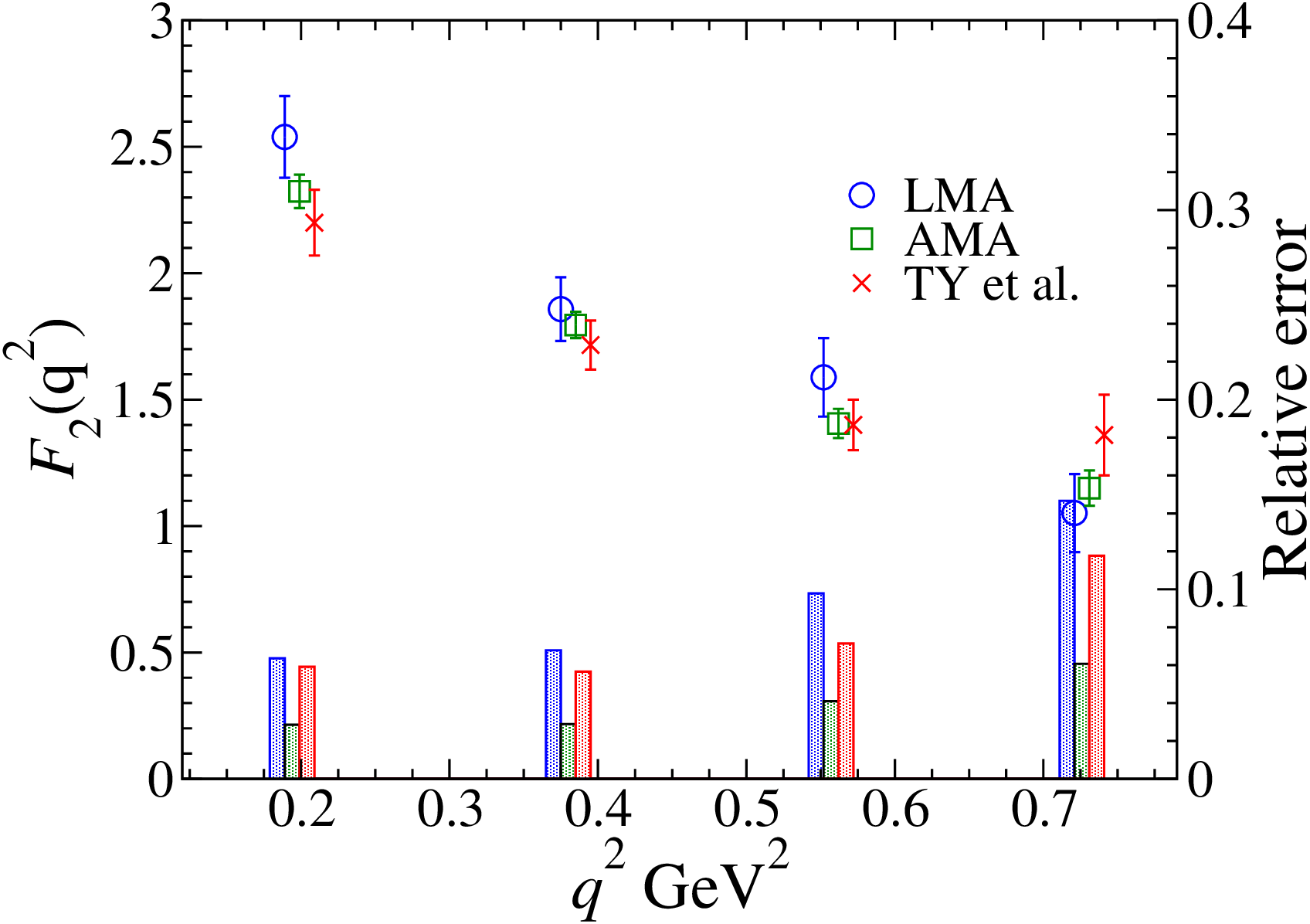}
\caption{Isovector form factor $F_1(q^2)$ and $F_2(q^2)$ obtained in LMA/AMA 
and presented in TY et al.\cite{Yamazaki:2009zq}
at $m=0.005$. 
The bars denote the relative error for each $q^2$ results.}\label{fig:f12}
\end{center}
\end{figure}

To test AMA for more involved observables, we compare
the three-point functions from AMA/LMA with parameters 
shown in Table \ref{tab:param}
and the previous result 
in \cite{Yamazaki:2009zq}. 
Nucleon isovector form factors, which are particularly  
feasible test bed, are extracted from the ratio of three and two point functions: 
\begin{equation}
  R_\mu(t_1,t,t_0|p_1,p_0) = K
  \frac{C^N_{J_\mu}(\vec q,t)}{C_G^N(t_1-t_0,0)}\bigg[ 
  \frac{C_L^N(t_1-t,\vec q)C_G^N(t-t_0,0)C_L^N(t_1-t_0,0)}
       {C_L^N(t_1-t,0)C_G^N(t-t_0,\vec q)C_L^N(t_1-t_0,\vec q)}\bigg]^{1/2}~~.
  \label{eq:r_mu}
\end{equation} 
Here  $C_{L,G}^N(t,\vec q)$ is the two-point function 
of nucleon with local (L) or Gaussian (G) sink at spatial momentum $\vec q$, 
$C^N_{J_\mu}(\vec q,t)$ is the three-point functions with 
vector current $J_\mu$, and $K=\sqrt{2(E_N+m_N)/E_N}$.
See \cite{Yamazaki:2009zq} for more details.
Isovector form factors extracted from Eq.(\ref{eq:r_mu}) are 
shown in Figure \ref{fig:f12}.
Precision of $F_{1,2}(q^2)$ in AMA is better than the 
previous results \cite{Yamazaki:2009zq} at all transfer momenta $q^2$ examined.
This demonstrates that AMA is effective to
reduce errors for many lattice observables.

\section{Summary}
In this proceedings, we have shown several results using the new class of 
(bias-free) error reduction techniques 
called as covariant approximation averaging (CAA). 
We proposed all-mode-averaging (AMA), 
in which contributions from {\it all} eigen modes are 
taken into account in the averaged approximation 
with the sloppy CG using deflation.
AMA is applicable to a broad variety of observables including nucleon spectrum,
three-point functions and other composite correlators. 
We compared the nucleon mass and isovector form factors 
for realistic $N_f=2+1$ DWF configurations with lattice volume 
($2.7$ fm$^3$) and light quark masses ($m_\pi\simeq 0.3$--0.4 GeV).
Significant gain in computational cost, 
up to 15 times is observed when compared with traditional many-source method 
without the deflation (and 4 times compared with the case with deflation). 
AMA significantly reduces error for cases in which LMA can not so much. 
Using this technique, calculations of the nucleon 
electric dipole moment and the hadronic contributions to 
the muon g-2 are now underway.

Numerical calculations were performed using the RICC at RIKEN and the Ds cluster at FNAL. 
This work was supported by the Japanese Ministry of Education Grant-in-Aid, 
Nos. 22540301 (TI), 23105714 (ES), 23105715 (TI) and U.S. DOE grants DE-AC02-98CH10886 (TI) 
and DE-FG02-92ER40716 (TB).
We also thank BNL, the RIKEN BNL Research Center, and USQCD for providing resources 
necessary for completion of this work.

\end{document}